\documentstyle[aps,mnb]{revtex}

\ifx\ps@drafts\undefined\else
\fi

\newif\ifIncludeFigs\IncludeFigsfalse
\newif\ifTwoUp\TwoUpfalse
\newif\ifEPSF\EPSFfalse
\IncludeFigstrue

\ifTwoUp 
  \twoup
\fi

\ifIncludeFigs
   \ifEPSF 
      \@input epsf %
      \ifx\epsfbox\undefined
         \EPSFfalse
	 \IncludeFigsfalse
      \else
      \fi
  \else  
  \fi
\else
\fi

\ifx\endfloat\notincluded\else
  \nomarkersintext
\fi

\catcode`@=11

\ifx\@ove@rcfont\undefined 
  \ifx\citen\undefined
     \def\citen#1{\begingroup \def\@cite##1##2{{##1}}%
	\@citex[]{#1}\endgroup}
  \fi
\else
  \def\@cite#1{$\@ove@rcfont\m@th^{[{#1}]}$}
\fi

\let\mdef\def

\mdef\tableline{\hline}  

\newdimen\slashraise \slashraise=0.33pt
\mathchardef\fslash="0236

\def\slash@char#1#2{%
   \setbox0=\hbox{$\m@th#2$}
   \dimen0=\wd0                                 
   \dimen2=-\dp0 \advance\dimen2 by \slashraise
   \setbox1=\hbox{$\m@th#1\mkern-13mu\fslash$}
	 \dimen1=\wd1               
   \ifdim\dimen0>\dimen1                        
      \rlap{\hbox to \dimen0{\hss\raise\dimen2\box1\hss}}%
      #2                          		
   \else                                        
      \rlap{\hbox to \dimen1{\hss\box0\hss}}    
      \raise\dimen2\box1                              
   \fi}                                         %

\def\slashchar#1{\mathpalette\slash@char#1}

\def\vereq#1#2{\lower3pt\vbox{\baselineskip1.5pt \lineskip1.5pt
\ialign{$\m@th#1\hfill##\hfil$\crcr#2\crcr\sim\crcr}}}

\mdef\eqb{\begin{equation}}
\mdef\eqe{\end{equation}}

\mdef\MeV{{\rm \,MeV}}
\mdef\GeV{{\rm \,GeV}}

\mdef\ie{{\it i.e.}}
\mdef\etal{{\it et al.}}

\mdef\Or#1{O(#1)} 
\mdef\Order{O}

\mdef\BRA#1{\left\langle #1\right|}
\mdef\bra#1{\langle #1 |}
\mdef\KET#1{\left| #1\right\rangle}
\mdef\ket#1{| #1\rangle}
\mdef\VEV#1{\left\langle #1\right\rangle}
\mdef\vev#1{\langle 0 | #1 | 0\rangle}
\mdef\braket#1#2{\langle#1 | #2\rangle}

\mdef\Tr{\mathop{\rm Tr}\nolimits}
\mdef\tr{\mathop{\rm tr}\nolimits}
\mdef\str{\mathop{\rm str}\nolimits}
\mdef\sdet{\mathop{\rm sdet}\nolimits}
\mdef\det{\mathop{\rm det}\nolimits}
\mdef\Str{\mathop{\rm Str}\nolimits}
\mdef\Sdet{\mathop{\rm Sdet}\nolimits}
\mdef\Det{\mathop{\rm Det}\nolimits}
\mdef\Im{\mathop{\rm Im}\nolimits}
\mdef\trd{\mathop{\rm tr}_D\nolimits}

\mdef\scr#1{{\cal #1}}
\mdef\scrD{\scr{D}}
\def\Square{\mathchoice{\square{6pt}}{\square{5pt}}{\square{4pt}}
    {\square{3pt}}}
\def\square#1{\mathop{\mkern0.5\thinmuskip\vbox{\hrule\hbox{\vrule
    \hskip#1 \vrule height#1 width 0pt \vrule}\hrule}\mkern0.5\thinmuskip}}
\mdef\lr{\leftrightarrow}

\mdef\scr#1{{\cal #1}}

\mdef\chisim{$SU(3)_L \times SU(3)_R$}
\mdef\im{{\rm i}}
\mdef\pintegral{\int\!{d^D\,p\over(2\pi)^D}}

\mdef\L#1{\,\mu_{#1}}
\mdef\Lq{\,\mu_q}
\mdef\LM{\L{d}}

\mdef\onpif#1{{#1 \over 16 \pi^2 f^2}}

\mdef\qt{{\tilde q}}
\mdef\qbar{{\overline q}}
\mdef\qtbar{{\overline\qt}}
\mdef\phit{{\tilde\phi}}

\mdef\Ht{{\tilde H}}
\mdef\Bt{{\tilde B}}
\mdef\etat{{\tilde \eta}}
\mdef\epshat{\hat\epsilon}

\preprint{
${\rm JHU-TIPAC-98005} \atop{{\rm hep-th/9803113}}$}

\begin{document}
\tightenlines
\begin{titlepage}
\title{HeatK: A Mathematica Program for Computing Heat Kernel Coefficients}

\author{Michael J.~Booth%
\footnote{Electronic Address: booth@bohr.pha.jhu.edu}
}
\address{%
Department of Physics and Astronomy \\
The Johns Hopkins University \\
Baltimore, MD 21218
}

\maketitle
\begin{abstract}
Heat kernel coefficients encode the short distance behavior of
propagators in the presence of background fields, and are thus useful
in quantum field theory.  We present a Mathematica program for
computing these coefficients and their derivatives, based on an
algorithm by Avramidi\cite{Avramidi:npb91}.

\end{abstract}

\end{titlepage}
\def\thesection{\arabic{section}}
\def\thesubsection{\arabic{section}.\arabic{subsection}}

\section{Program Summary}

\noindent
\emph{Title of program:} 
HeatK
\\[\baselineskip]
\noindent
\emph{Version number:} 
1.0
\\[\baselineskip]
\noindent
\emph{Available at:} 
{\tt http://fermi.pha.jhu.edu/\~{}booth/HeatK/ }
\\[\baselineskip]
\noindent
\emph{Programming Language:} 
{\it Mathematica} 2.2 or higher
\\[\baselineskip]
\noindent
\emph{Platform:} 
Any platform supporting {\it Mathematica} 2.2
\\[\baselineskip]
\noindent
\emph{Keywords:} 
Heat kernel, HDMS coefficients, Effective Action
\\[\baselineskip]
\noindent
\emph{Nature of physical problem:} 
Computation of heat kernel coefficients.
\\[\baselineskip]
\noindent
\emph{Method of solution:} 
The non-recursive solution given by Avramidi \cite{Avramidi:npb91}
\\ \\ 
\noindent
\emph{Limitations:} 
Degree of the heat kernel coefficient.

\section{Introduction}
Heat Kernel (also know as HDMS or Seeley-de Witt) coefficients have
many applications in physics and mathematics ( a comprehensive list
can be found in a paper by Avramidi and
Schwimming\cite{Avramidi:1995}).
Because of this interest,
there is an extensive literature describing their computation and
properties.  
Unfortunately for those who are primarily interested in using
the coefficients as inputs to a calculation, there is not one
comprehensive source for them, and moreover a number of different sign
and normalization conventions are in use, making it difficult to
combine sources.  A further complication is that many applications
require derivatives of the coefficients, of which only a few
can be found in the literature.
HeatK was created to resolve these problems and provide an environment
which allows easy manipulation of the results.

\section{Theory}

Consider a general manifold of dimension $d$ with metric $g_{\mu\nu}$,
fields $\varphi^{(i)}$ and derivative $\nabla_\mu$ acting 
such that 
$[\nabla_\mu,\nabla_\nu]\varphi$ = $F_{\mu\nu} \varphi$ and 
$[\nabla_\mu,\nabla_\nu]\varphi^\sigma$ = 
$R^\sigma{}_{\tau\mu\nu}\varphi^\tau$.  For a differential
operator $H = -\Square + Q$, with $\Square = g_{\nu\mu}\nabla^\nu\nabla^\mu$
and $Q$ a matrix-valued potential, 
the heat kernel expansion	
is developed by considering the 
exponential $U(t) = \exp(-t H)$, which satisfies the equation
\eqb
\label{eqn:heat}
 {\partial U(t) \over \partial t}  = - H U(t).
\eqe
The kernel of the operator $U$ is then parameterized as%
\footnote{We follow the conventions of Ref.\citen{Avramidi:npb91} throughout.}
\eqb
\label{eqn:asymp}
U(t|x,x') = {1 \over (4\pi t)^{d/2}}\Delta^{1/2}
e^{-\sigma/ 2t}
	\sum\limits_{k\ge 0}{(-t)^k \over k!} a_k(x,x'),
\eqe
where $a_k$ are the heat kernel coefficients, 
$\sigma(x,x')$ is half the square of the geodesic distance between the
points $x$ and $x'$ and
\eqb
\Delta(x,x') = g^{-1/2}(x)
\det\left(-\nabla_{\mu'}\nabla_\nu\sigma(x,x')\right)g^{-1/2}(x').
\eqe
is the Van Fleck-Morette determinant.

%
To obtain expressions for the $a_k$, we make use of the non-recursive
algorithm of Avramidi\cite{Avramidi:npb91,Avramidi:thesis}.  
Here we content ourselves
with only a brief outline of the method, and refer the reader to the
original papers for a more detailed description.

Given the above data, 
we can then construct the two tangent vectors, 
$\sigma_\nu = \nabla_\nu \sigma$ and $\sigma_{\nu'} = \nabla_{\nu'} \sigma$,
noting that 
${1\over2}\sigma_\nu\sigma^\nu = \sigma = {1\over2}\sigma_{\nu'}\sigma^{\nu'}$.
We also have the
coincidence limits
\eqb
[\sigma]=[\sigma^\mu]=[\sigma^{\mu'}]=0,
\eqe
with 
$$
[f(x,x')]\equiv\lim_{x\to x'} f(x,x'),
$$
It is also natural to define the tensor 
$\eta^{\mu'}{}_{\nu} = \nabla_{\nu} \sigma^{\mu'}$ and its inverse
$\gamma^{\mu}{}_{\nu'}$.
We can then define the parallel displacement operator $\scr{P}(x,x')$ as
the solution to $\sigma_\mu \nabla^\mu \scr{P} = 0$ with $[\scr{P}] = 1$
(When acting on vector indices, it is conventional to write
$g^\mu{}_{\nu'}$ instead of $\scr{P}^\mu{}_{\nu'}$).
In what follows it is convenient to work with quantities bases at $x'$, 
so we define
$\bar\nabla_{\nu'} = \gamma^\mu{}_{\nu'} \nabla_{\mu}$ and the ``gauge''
field $\bar{\scr{A}}_{\mu'} = \scr{P}^{-1}\bar\nabla_{\mu'} \scr{P}$. 

Introduce a functional basis 
\begin{eqnarray}
\vert 0\rangle &\equiv& 1, \\
\vert n\rangle & \equiv& \vert \nu'_1...\nu'_n\rangle
={(-1)^n\over n!}\sigma^{\nu'_1}\cdots
\sigma^{\nu'_n}, \qquad (n\ge 1),
\end{eqnarray}
its dual,
\eqb
\langle m\vert\equiv\langle\mu'_1\cdots\mu'_m\vert=(-1)^m g^{\mu_1}_{\mu'_1}\cdots
g^{\mu_m}_{\mu'_m}\nabla_{(\mu_1}\cdots\nabla_{\mu_m)}\delta (x,x'),
\eqe
and the scalar product
\eqb
\langle m|n\rangle=\int\,d^n x\,\langle\mu'_1\cdots\mu'_m\vert\nu'_1...\nu'_n\rangle,
\eqe
so that 
\eqb
\langle m\vert n\rangle=\delta_{mn}\, \delta^{\nu_1\cdots\nu_n}_{\mu_1\cdots\mu_n}
= \delta_{mn}\,\delta^{\nu_1}_{(\mu_1}\cdots\delta^{\nu_n}_{\mu_n)}.
\eqe
For a field $\varphi$ we then have
\eqb
\langle m\vert\varphi\rangle
=\left[\nabla_{(\mu_1}\cdots\nabla_{\mu_m)}\varphi\right],
\eqe
and the covariant Taylor series is simply 
\eqb 
\vert\varphi\rangle={\cal P}\sum_{n\ge 0}\vert n\rangle\langle n\vert\varphi\rangle.
\eqe
The advantage of this machinery is that it allows us to take the
familiar HDMS recursion relation:
\eqb
\left(1+{1\over k}\sigma^\mu\nabla_\mu\right)a_k = 
	\Delta^{-1/2} H (\Delta^{1/2} a_{k-1}) \equiv F a_{k-1},
\eqe
which follows from inserting the expansion Eq.~(\ref{eqn:asymp})
into Eq.~(\ref{eqn:heat}) 
and give meaning to its formal solution 
\eqb
a_k=\scr{P}\left(1+{1\over k}D\right)^{-1}\bar F
\left(1+{1\over k-1}D\right)^{-1} \bar F\cdots(1+D)^{-1} \bar F,
\eqe
(where $D\equiv \sigma \cdot \nabla$ and 
$\bar F \equiv \scr{P}^{-1} F \scr{P}$).
This is accomplished by noting that $D|n\rangle = n |n\rangle$, so that
\eqb
\left(1+{{1}\over {k}}D\right)^{-1}
=\sum_{n\ge 0}{k\over {k+n}}\vert n\rangle\langle n\vert,
\eqe
from which it follows that
\begin{eqnarray}
\lefteqn{\langle n\vert a_k\rangle
=\sum_{n_1,\cdots,n_{k-1}\ge 0}{k\over k+n}\cdot
{k-1\over k-1+n_{k-1}}\cdots{1\over 1+n_1}}
\qquad\ \,\nonumber\\[10pt]
& &\times \langle n\vert \bar F\vert n_{k-1}\rangle 
	\langle n_{k-1}\vert \bar F\vert n_{k-2}\rangle\cdots
	\langle n_1\vert \bar F\vert 0\rangle,
\end{eqnarray}
with 
\begin{eqnarray}
\langle m\vert \bar F\vert n\rangle
&=&\left[\nabla_{(\mu_1}\cdots \nabla_{\mu_m)}\bar F{{(-1)^n}\over {n!}}
\sigma^{\nu'_1}\cdots \sigma^{\nu'_n}\right] \nonumber\\
&=&{m\choose n}
\delta^{\nu_1\cdots\nu_n}_{(\mu_1\cdots\mu_n}Z_{\mu_{n+1}\cdots\mu_m)}
-{m\choose n-1}\delta^{(\nu_1\cdots\nu_{n-1}}_{(\mu_1\cdots\mu_{n-1}}
Y^{\nu_n)}_{\ \ \ \mu_n\cdots \mu_m)}
\nonumber\\[10pt]
& &
+{m\choose n-2}\hat 1\delta^{(\nu_1\cdots\nu_{n-2}}_{(\mu_1\cdots\mu_{n-2}}
X^{\nu_{n-1}\nu_n)}_{\ \ \ \ \ \ \ \ \mu_{n-1}\cdots\mu_m)}
\end{eqnarray}
where 
$$
Z_{\mu_1\cdots\mu_n}
=(-1)^n\left[\bar\nabla_{(\mu'_1}\cdots\bar\nabla_{\mu'_n)}Z\right],
$$
$$
Y^{\nu}_{\ \ \mu_1\cdots\mu_n}
=(-1)^n\left[\bar\nabla_{(\mu'_1}\cdots\bar\nabla_{\mu'_n)}Y^{\nu'}\right],
$$
$$
X^{\nu_1\nu_2}_{\ \ \ \ \mu_1\cdots\mu_n}
=(-1)^n\left[\bar\nabla_{(\mu'_1}\cdots\bar\nabla_{\mu'_n)}
	X^{\nu'_1\nu'_2}\right].
$$
Here $X$, $Y$ and $Z$ are simply the coefficients of the Hamiltonian operator
transported to the point $x'$, that is 
\begin{eqnarray}
\bar F 
  &=&{\scr{P}^{-1}(\Delta^{1/2}\Square\Delta^{-1/2} + Q) \scr{P}} \nonumber\\
&=& {\cal P}^{-1}\Delta^{1/2}\bar\nabla_{\mu'}
\Delta^{-1}X^{\mu'\nu'}\bar\nabla_{\nu'}
\Delta^{1/2}{\cal P} + \bar Q
\nonumber\\[10pt]
&=&  X^{\mu'\nu'}\bar\nabla_{\mu'}\bar\nabla_{\nu'}
+ Y^{\mu'}\bar\nabla_{\mu'}+Z.
\end{eqnarray}

\subsection{Derivatives}
Avramidi's formalism is naturally adapted to computing the symmetrized
derivatives of the $a_k$, but frequently one requires the un-symmetrized
derivatives.  In order to obtain these, we can continue in the spirit
of Avramidi and introduce matrix elements for the derivative operators.
Defining
\eqb
T^{\nu_1'\ldots\nu_n'}_{\mu_1\ldots\mu_m}[h_{\lambda_1\ldots\lambda_n}] = 
  \langle 0|\nabla_{\mu_1}\ldots\nabla_{\mu_m} \scr{P} 
	h_{\lambda_1\ldots\lambda_n} |n\rangle,
\eqe
we then have
\eqb
[\nabla_{\nu_1}\ldots\nabla_{\nu_n} a_k] = 
\sum_m T^{\mu_1\ldots\mu_m}_{\mu_1 \ldots \mu_n}[1] \langle m|a_k\rangle.
\eqe
Noting that
\eqb
\nabla_{\mu} \scr{P} h_{\lambda_1\ldots\lambda_l} |n\rangle = 
  -\scr{P} h_{\lambda_1\ldots\lambda_n}\eta^{(\nu_1'}_{\mu}
    |n-1\rangle^{\nu_2\ldots\nu_n)} + 
  \scr{P}\eta^{\nu'}{}_{\mu}( \bar{\scr{A}}_{\nu'}h_{\lambda_1\ldots\lambda_n} 
 +	\sum_{i} \bar{\scr{B}}^{\sigma}{}_{\lambda_i\nu'} 
    h_{\lambda_1\ldots\sigma\ldots\lambda_n}) |n\rangle,
\eqe
where $\scr{B}$ is a special case of $\scr{A}$: 
$\scr{B}_{\alpha'}{}^{\nu'}{}_\mu \equiv g^{\sigma}{}_{\alpha'} \nabla_{\mu} 
g^{\nu'}{}_{\sigma}$,
it is possible to derive a recursive expression for $T$:
\begin{eqnarray}
T^{\nu_1'\ldots\nu_n'}_{\mu_1\ldots\mu_m}[h_{\lambda_1\ldots\lambda_l}] &=& 
 -\sum_{k=0}^{m-n-1} 
  T^{(\nu_2'\ldots\nu_n'|\kappa_1'\ldots\kappa_k'|}_{\mu_1\ldots\mu_{m-1}}
[\eta^{\nu_1')}{}_{\mu_m(\kappa_1'\ldots\kappa_k)'} 
	h_{\lambda_1\ldots\lambda_l}]
 \nonumber\\
&&+
 \sum_{k=1}^{m-n-1} 
	T^{\nu_1'\ldots\nu_n'\kappa_1'\ldots\kappa_k'}_{\mu_1\ldots\mu_{m-1}}
  [\scr{A}_{\mu_m(\kappa_1'\ldots\kappa_k')} h_{\lambda_1\ldots\lambda_l}]
\nonumber \\
&&+
\sum_{k=1}^{m-n-1}\sum_{i=1}^{l} 
	T^{\nu_1'\ldots\nu_n'\kappa_1'\ldots\kappa_k'}_{\mu_1\ldots\mu_{m-1}}
[\scr{B}_{\lambda_i}{}^\sigma{}_{\mu_m(\kappa_1'\ldots\kappa_k')} 
	h_{\lambda_1\ldots\sigma\ldots\lambda_l}].
\end{eqnarray}

\section{Program Description}
\subsection{Functions}
The program implements most of the above functions, with the following
correspondence:
\begin{eqnarray}
{[a_k]} &=& {\tt DCL[k]}, \nonumber\\
{[\nabla_{\mu_1} \ldots \nabla_{\mu_n} a_k]} &=& 
	{\tt DCL[k, \mu_1, \ldots, \mu_n]}, \nonumber\\
{X^{\alpha\beta}_{(\nu_1 \ldots \nu_n)} } &=& 
	{\tt XX[\alpha, \beta, \{\nu_1, \ldots, \nu_n\}]}, \nonumber\\
{Y^{\alpha}_{(\nu_1 \ldots \nu_n)} } &=& 
	{\tt YY[\alpha, \{\nu_1, \ldots, \nu_n\}]}, \nonumber\\
{Z_{(\nu_1 \ldots \nu_n)} } &=& 
	{\tt ZZ[\{\nu_1, \ldots, \nu_n\}]}, \nonumber\\
{[\nabla_{(\mu_1} \ldots \nabla_{\mu_n)} Q] } &=& 
	{\tt QQ[\{\mu_1, \ldots, \mu_n\}] }, \nonumber\\
{[Q]} &=& {\tt QQ[\ ] }, \nonumber\\
{F^{\mu\nu}} &=& {\tt F[\mu, \nu] },  \nonumber\\
{R^{\mu\nu\alpha\beta}} &=& {\tt R[\mu, \nu, \alpha, \beta] }.
\end{eqnarray}
In general, as already seen above, we use a list of bracketed indices
to indicate symmetrized differentiation, so we also have for example
\eqb
{\nabla_{(\mu_1} \ldots \nabla_{\mu_n)} F^{\alpha\beta} } = 
 	F^{\alpha\beta}{}_{(\nu_1 \ldots \nu_n)} = 
	{\tt F[\alpha, \beta, \{\mu_1, \ldots, \mu_n\}] }.
\eqe
For completeness, we also list the following functions, although they
are essentially internal functions:
\begin{eqnarray}
{\langle m\vert F\vert n \rangle} &=& 
	{\tt MF[\{\nu_1,\ldots,\nu_n\}, \{\mu_1,\ldots,\mu_m\}]  }, \nonumber\\
{T^{\nu_1\ldots\nu_n}_{\mu_1\ldots\mu_m}[h_{\lambda_1\ldots\lambda_n}]} &=& 
	{\tt MD[\{\nu_1,\ldots,\nu_n\}, 
		\{\mu_1,\ldots,\mu_m\}, \{\lambda_1\ldots\lambda_n\}, 
		h_{\lambda_1\ldots\lambda_n}]},
\end{eqnarray}
where we note that contrary to the convention, ${\tt MD}$ is not
symmetric in the first set of indices.

\subsection{Options}
For the primary function {\tt DCL} there
are several options which control the normalization and degree
of simplification.  They are:
\begin{description}
\item[{\tt HeatKNorm}] \mbox{}\\
  This controls the normalization that is used for the heat kernel
  coefficients returned by {\tt DCL}. The default is 
  {\tt Avramidi}, which naturally invokes Avramidi's
  normalization $(-1)^k/k!$. It can also be set to a function which
  should expect two arguments: the degree of the coefficient, and the
  number of derivatives.

\item[{\tt EvaluateK}] \mbox{}\\
	The curvature tensor occurs frequently through
	  its symmetrized derivative 
$K^\alpha{}_\beta{}_{(n)} = \nabla_{(\nu_1\ldots \nu_{n-2}}
	R^\alpha{}_{\nu_{n-1}|\beta|\nu_n)}$.  Setting this option to
{\tt True} expands $K$ in terms of $R$.
\item[{\tt EvaluateSymmetricDerivs}] \mbox{}\\
  Setting this to {\tt True} means that objects such as
	${\tt Q[\{\nu_1,\ldots,\nu_n\}] }$ will be expanded in terms
	of explicitly symmetrized derivatives (represented by the operator
	{\tt Dx}), so that for example, ${\tt Q[\{\mu_1, \mu_2\}] = 
	(Dx[Q[\ ], \mu_1, \mu_2] + Dx[Q[\ ], \mu_2, \mu_1])/2}$.

\item[{\tt DoSimplifications }] \mbox{}\\
  This option controls whether derivatives will be re-ordered and other 
  simplifications performed.  

\item[{\tt DoBasicSimplifications }] \mbox{}\\
  This option controls whether to do basic simplifications like
  contracting indices. Setting this to {\tt False} is
  really only useful when debugging.
\end{description}

There is also a global option defined for the {\tt HeatK}:
\begin{description}
\item[{\tt EvaluateZZ}]  \mbox{}\\
  Setting this option to {\tt False} means that 
{\tt ZZ}, {\tt YY} and {\tt XX} will
not be evaluated.  This makes the results much more compact, though cryptic.
\end{description}

Two variables which describe the form of the derivative
operator.  The first, {\tt \$HeatKMetricStyle} indicates whether one
wishes to use the GR style metric $diag(-1,1,\dots,1)$ or the HEP
style metric $diag(1,-1,\ldots,-1)$.  Setting it to $-1$ (the default)
selects the former.  The second variable, {\tt \$ScalarCurvatureFactor}
is used to include an explicit factor of the scalar curvature ($R$) into
the Hamiltonian.  For example, with ${\tt \$HeatKMetricStyle}=-1$ and
${\tt \$ScalarCurvatureFactor}=-1/6$, one has
$H = -\Square + Q - R/6$.

\subsection{Examples}


The following presents a simple example illustrating the use of the package,
and showing how to change the settings to conform to various conventions.
First load the package, assuming it is in the search path:
\beginmath
<<HeatK.m
\endmath
By default, the package is set up to follow the sign and normalization
conventions of Avramidi's article \cite{Avramidi:npb91}
We can check that it reproduces those results.

\beginmath
a1 = DCL[1]
\endmath

\beginmathout
ZZ
  ;()
\endmathout

\beginmath
a2 = DCL[2] // Expand
\endmath

\beginmathout
                 ZZ
                   ;(u1,u1)
ZZ    ** ZZ    - ----------
  ;()      ;()       3
\endmathout

\beginmath
a3 = DCL[3] // Expand
\endmath

\beginmathout
-YY        ** ZZ           YY           ** ZZ
   u1;(u2)      ;(u1,u2)     u1;(u2,u2)      ;(u1)
------------------------ - ----------------------- - 
           3                          4
 
  ZZ    ** ZZ           ZZ      ** ZZ
    ;()      ;(u1,u1)     ;(u1)      ;(u1)
  ------------------- - ------------------ - 
           2                    2
 
  ZZ         ** ZZ
    ;(u1,u1)      ;()
  ------------------- + ZZ    ** ZZ    ** ZZ    + 
           2              ;()      ;()      ;()
 
  XX              ZZ           ZZ
    u1,u2;(u3,u3)   ;(u1,u2)     ;(u1,u1,u2,u2)
  -------------------------- + ----------------
              6                       10
\endmathout
It is not difficult to see that the above results agree with those
presented by Avramidi. Now, if we prefer, we can follow the
conventions in his thesis \cite{Avramidi:thesis}.

\beginmath
$HeatKMetricStyle = -1;
\endmath

\beginmath
a1 = DCL[1]
\endmath

\beginmathout
ZZ
  ;()
\endmathout

\beginmath
a2 = DCL[2] // Expand
\endmath

\beginmathout
                 ZZ
                   ;(u1,u1)
ZZ    ** ZZ    + ----------
  ;()      ;()       3
\endmathout

\beginmath
a3 = DCL[3] // Expand
\endmath

\beginmathout
-YY        ** ZZ           YY           ** ZZ
   u1;(u2)      ;(u1,u2)     u1;(u2,u2)      ;(u1)
------------------------ - ----------------------- + 
           3                          4
 
  ZZ    ** ZZ           ZZ      ** ZZ
    ;()      ;(u1,u1)     ;(u1)      ;(u1)
  ------------------- + ------------------ + 
           2                    2
 
  ZZ         ** ZZ
    ;(u1,u1)      ;()
  ------------------- + ZZ    ** ZZ    ** ZZ    + 
           2              ;()      ;()      ;()
 
  XX              ZZ           ZZ
    u1,u2;(u3,u3)   ;(u1,u2)     ;(u1,u1,u2,u2)
  -------------------------- + ----------------
              6                       10
\endmathout

Now let's now back up and see what some of these expressions are in
more familiar notation.  We can do this by changing the options for
HeatK, allowing {\tt XX}, {\tt YY} and {\tt ZZ} to be evaluated.  We
also switch back to the conventions in the nuclear physics article
\cite{Avramidi:npb91}.

\beginmath
SetOptions[HeatK, EvaluateZZ -> True];
$HeatKMetricStyle = -1;
\endmath

\beginmath
ZZ[{}] // ContractAll
\endmath

\beginmathout
     R
QQ - -
     6
\endmathout

{\tt SimplifyCurvature} uses some tricks to put the curvature tensor into a
{\tt "}canonical{\tt "} form, while {\tt ContractAll} and 
{\tt ReduceContractions}
perform contractions, and simplify them - {\tt ZZ}, {\tt XX} and 
{\tt YY} do not
simplify their results , because they are usually not called directly.

\beginmath
ZZ[{u1, u1}] // ContractAll // SimplifyCurvature // 
	ReduceContractions
\endmath

\beginmathout
-F      ** F                     R
  u1,u2     u1,u2                 ;(u1,u1)
----------------- + QQ         - --------- + 
        2             ;(u1,u1)       5
 
                        2               2
  F      R        R         R
   u1,u2  u1,u2    u1,u2     u1,u2,u3,u4
  ------------- + ------- - -------------
        3           30           30
\endmathout
We see that these results agree with Eqns. 3.83a and 3.83b of 
Ref.\citen{Avramidi:npb91}.

Now let's compare with a standard reference, Barvinskii and
Vilkoviskii's (BV) Physics Reports \cite{Barvinskii:physrep85}.  
Avramidi defines his heat
kernel coefficients with an explicit weight of $1/k!$, so we have to
divide by $1/k!$ to get the more standard coefficients used by BV.
Also, BV uses ($is$) rather than $t$ as the exponential parameter
introduces an extra factor of $(-1)^k$ in their definition of the
coefficients.  All told, their conventions correspond to the
following:
\beginmath
SetOptions[DCL, HeatKNorm -> ((-1)^(#1)/(#1)!&)];
$HeatKMetricStyle = -1; $ScalarCurvatureFactor = 1/6;
\endmath
We must also note that their matrix $P$ is opposite in sign to
Avramidi's {\tt QQ}, so we correct for that as well.

\beginmath
DCL[1] /. _QQ -> -QQ
\endmath

\beginmathout
QQ
\endmathout

\beginmath
(a2 = DCL[2] // Expand) /. _QQ -> -QQ
\endmath

\beginmathout
D  D  (QQ)   D  D  (R)              F      ** F
 u1 u1        u1 u1      QQ ** QQ    u1,u2     u1,u2
---------- + --------- + -------- + ---------------- - 
    6           180         2              12
 
                        2               2
  F      R        R         R
   u1,u2  u1,u2    u1,u2     u1,u2,u3,u4
  ------------- - ------- + -------------
       36           180          180
\endmathout

\beginmath
DCL[1, u1] /. _QQ -> -QQ
\endmath

\beginmathout
D  (QQ)   D  (F     )
 u1        u2  u1,u2
------- + -----------
   2           6
\endmathout

\beginmath
DCL[1, u1, u2] /. _QQ -> -QQ
\endmath

\beginmathout
D  D  (QQ)   D  D  (F     )   D  D  (F     )
 u1 u2        u2 u3  u1,u3     u1 u3  u2,u3
---------- + -------------- + -------------- - 
    3              12               12
 
  D  D  (R)   D  D  (R     )   QQ ** F
   u1 u2       u3 u3  u1,u2           u1,u2
  --------- + -------------- + ------------ + 
     180            60              6
 
  F      ** QQ   F      ** F        F      ** F
   u1,u2          u1,u3     u2,u3    u2,u3     u1,u3
  ------------ + ---------------- + ---------------- - 
       3                12                 12
 
  R      R        F      R
   u1,u3  u2,u3    u3,u4  u1,u3,u2,u4
  ------------- + ------------------- + 
       45                 18
 
  R      R              F      R
   u3,u4  u1,u3,u2,u4    u3,u4  u1,u4,u2,u3
  ------------------- + ------------------- + 
          180                   18
 
  R      R              R            R
   u3,u4  u1,u4,u2,u3    u1,u3,u4,u5  u2,u3,u4,u5
  ------------------- + -------------------------
          180                      90
\endmathout

More Examples (as Mathematica notebooks) can be found at the WWW-site
listed in the program summary.

\acknowledgements
I would like to thank R.~G.~Sachs for encouraging me to present this
work.  This work was supported by the National Science Foundation under
Grant PHY-940457.

\mdef\jvp#1#2#3#4{#1~{\bf #2}, #3 (#4)}
\mdef\PR#1#2#3{\jvp{Phys.~Rev.}{#1}{#2}{#3}}
\mdef\PRD#1#2#3{\jvp{Phys.~Rev.~D}{#1}{#2}{#3}}
\mdef\PRL#1#2#3{\jvp{Phys.~Rev.~Lett.}{#1}{#2}{#3}}
\mdef\PLB#1#2#3{\jvp{Phys.~Lett.~B}{#1}{#2}{#3}}
\mdef\NPB#1#2#3{\jvp{Nucl.~Phys.~B}{#1}{#2}{#3}}
\mdef\SJNP#1#2#3{\jvp{Sov.~J.~Nucl.~Phys.}{#1}{#2}{#3}}
\mdef\AP#1#2#3{\jvp{Ann.~Phys.}{#1}{#2}{#3}}
\mdef\PL#1#2#3{\jvp{Phys.~Lett.}{#1}{#2}{#3}}
\mdef\NuovoC#1#2#3{\jvp{Nuovo.~Cim.}{#1}{#2}{#3}}
\mdef\NPBPS#1#2#3{\jvp{Nucl.~Phys.~B~(Proc.~Suppl.)}{#1}{#2}{#3}}
\mdef\Prog#1#2#3{\jvp{Prog.~Theor.~Phys.}{#1}{#2}{#3}}
\mdef\ZPC#1#2#3{\jvp{Z.~Phys.~C}{#1}{#2}{#3}}

\bibliography{heatk}
\bibliographystyle{h-physrev}

  
\end{document}